\documentclass[reprint,aps,nofootinbib,twocolumn,amsmath,amssymb]{revtex4-1}
\usepackage{dcolumn}
\usepackage{bm}
\usepackage{graphicx}

\begin{document}

\title{Strong cosmic censorship in de Sitter space: As strong as ever}

\author{Hongbao Zhang$^{1,2}$}
\thanks{hzhang@vub.ac.be}
 \author{Zhen Zhong$^{1}$}
 \thanks{zhenzhong@mail.bnu.edu.cn}

\affiliation{
$^1$ Department of Physics, Beijing Normal University, Beijing 100875, China\\
 $^2$ Theoretische Natuurkunde, Vrije Universiteit Brussel,
 and The International Solvay Institutes, Pleinlaan 2, B-1050 Brussels, Belgium
}

\begin{abstract}
The strong cosmic censorship has recently been put into question for the charged black holes in de Sitter space.
We have performed the full non-linear evolution of the massless charged scalar field minimally coupled to the Einstein-Maxwell system in de Sitter space, and found that the non-linear effect can restore the strong cosmic censorship, making it as strong as ever.

\end{abstract}

\maketitle
{\it Introduction}.---The strong cosmic censorship (SCC) protects general relativity from losing its predictive power within its domain of validity in the sense that the Cauchy horizons (CH) appearing in the solutions to Einstein equation are believed to be unstable under generic perturbations and become inextendible singular boundaries.

The SCC keeps in good shape until the very recent observation of the violation in the highly charged Reissner-Nordstrom--de Sitter (RNdS) black hole, where the CH remains extendible against either the linear neutral scalar perturbation or the linear coupled electromagnetic and gravitational perturbations\cite{CCDHJ1,DRS1}. It is further shown in \cite{LZCCN} that the non-linear effect of these perturbations does not suffice to save the SCC out. However, different from the rotating Kerr-de Sitter black hole, where no violation of the SCC occurs\cite{DERS,Hod1}, a charged black hole, when formed, necessitates the presence of the remnant charged fields\cite{Hod2}. This has spurred further investigations of the fate of the SCC under the linear charged scalar perturbation\cite{Hod3,CCDHJ2,MTWZZ,DRS2}. As a result, the SCC is restored except in the highly near-extremal charged RNdS black hole, where the SCC can be violated  by the appropriately charged scalar field. This disturbing scenario has led one to rescue the SCC by allowing the initial data to be rough rather than smooth\cite{DRS1,DS}.

But nevertheless, note that the aforementioned violation regime of the SCC for the charged scalar field seems special in the sense that not only is the RNdS black hole is required to be highly near-extremal but also the scalar field is required to be appropriately charged. So before giving up on the smooth initial data, one has the last corner to clear up by examining whether the non-linear effect of the charged fields can save the SCC out of this special violation regime. To this end,  we investigate the full non-linear dynamics of the massless charged scalar field minimally coupled to the Einstein-Maxwell system in de Sitter space. As a result, we find that although the SCC can be violated at the linear level, the non-linear effect restores the SCC, making it as strong as ever even in de Sitter space. 

{\it Framework and methodology }.---We consider the massless charged scalar field minimally coupled to the Einstein-Maxwell system in de Sitter spacetime with the action
\begin{equation}
S=\int d^4x\sqrt{-g}[R-2\Lambda-F_{ab}F^{ab}-4(D_a\Psi)\overline{(D^a\Psi)}],
\end{equation}
where $R$ is the Ricci scalar, $\Lambda$ is the positive cosmological constant, $F_{ab}=\partial_aA_b-\partial_bA_a$ is the field strength with $A_a$ the electromagnetic potential, and $D_a=\nabla_a-iqA_a$ is the gauge covariant derivative with $q$ the charge of the scalar field $\Psi$.

We would like to focus on the spherically symmetric dynamics of the system in the double null coordinates with
\begin{eqnarray}
&&ds^2=-e^{\sigma(u,v)}dudv+r^2(u,v)d\Omega^2,\\
&&A=A(u,v)du, \quad \Psi=\Psi(u,v),
\end{eqnarray}
where $u$ and $v$ are the ingoing and outgoing coordinates, respectively. Accordingly, the equations of motion can be written as the evolution equations
\begin{eqnarray}
\label{re}0&=&r_{,uv}+r_{,u}r_{,v}/r+e^\sigma(1/r-\Lambda r)/4-e^{-\sigma}rA^2_{,v},\\
\label{sigmae}0&=&\sigma_{,uv}-2r_{,u}r_{,v}/r^2-e^\sigma/(2r^2)+4e^{-\sigma} A^2_{,v}+\nonumber\\
&&2iqA(\overline{\Psi}\Psi_{,v}-\Psi\overline{\Psi}_{,v})+2(\overline{\Psi}_{,u}\Psi_{,v}+\Psi_{,u}\overline{\Psi}_{,v})\\
\label{Ae}0&=&A_{,uv}+(2r_{,u}/r-\sigma_{,u})A_{,v}+\nonumber\\
&&e^\sigma[q^2A\Psi\overline{\Psi}
+iq(\overline{\Psi}\Psi_{,u}-\Psi\overline{\Psi}_{,u})/2],\\
\label{psie}0&=&\Psi_{,uv}+(r_{,u}\Psi_{,v}+r_{,v}\Psi_{,u})/r-\nonumber\\
&&iqA(\Psi_{,v}+r_{,v}\Psi/r)-iqA_{,v}\Psi/2,
\end{eqnarray}
supplemented by the constraint equations
\begin{eqnarray}
0&=&r_{,uu}-r_{,u}\sigma_{,u}+2r(\Psi_{,u}-iqA\Psi)(\overline{\Psi}_{,u}+iqA\overline{\Psi}),\label{c1}\\
0&=&r_{,vv}-r_{,v}\sigma_{,v}+2r\Psi_{,v}\overline{\Psi}_{,v},\label{c2}\\
0&=&(2r^2e^{-\sigma}A_{,v})_{,v}-iqr^2(\overline{\Psi}\Psi_{,v}-\Psi\overline{\Psi}_{,v}),\label{c3}
\end{eqnarray}
which will be preserved by the evolution equations in the whole domain of dependence due to the Bianchi identity provided that (\ref{c1}) is satisfied at some initial surface of constant $v$ as well as (\ref{c2}) and (\ref{c3}) satisfied at some initial surface of constant $u$\footnote{The roles of Eq.(\ref{Ae}) and Eq.(\ref{c3}) are actually interchangeable. As documented in the supplemental material, we shall take the former as the constraint equation to check the validity of our numerics.} .

It is noteworthy that not only are the above equations of motion invariant under the residual diffeomorphism
\begin{equation}
\sigma\rightarrow\sigma-\ln(\frac{du'}{du})-\ln(\frac{dv'}{dv}),\quad A\rightarrow A(\frac{du'}{du})^{-1},
\end{equation}
induced by the coordinate transformations of the form $(u,v)\rightarrow (u',v')$, but also under the following residual gauge transformation
\begin{equation}
A\rightarrow A+\frac{d\lambda}{du}, \quad \Psi\rightarrow e^{iq\lambda}\Psi
\end{equation}
with $\lambda$ the function of $u$. For later usage, we like to introduce the charge function and mass function as follows
\begin{eqnarray}
Q&=&2r^2e^{-\sigma}A_{,v},\\
M&=&(1+4e^{-\sigma}r_{,u}r_{,v})r/2+Q^2/(2r)-\Lambda r^3/6,
\end{eqnarray}
both of which are invariant under the above diffeomorphism and gauge transformation.

To model the fully non-linear dynamics of the RNdS black hole perturbed by the charged scalar field, we first take advantage of the above diffeomorphism and gauge freedom to specify the initial data as follows
\begin{equation}\label{initial}
 \Psi(0,v)=ae^{-(\frac{v-v_c}{w})^2}, \quad \Psi(u,0)=0, \quad A(u,0)=0
\end{equation}
together with the adaptive gauge choice for $\sigma$ at the initial double null surfaces. The initial data for $r$ as well as $A(0,v)$ can then be solved by the constraint equations with $r(0,0)=r_0$ sitting between the event and cosmological horizons of the initial RNdS black hole, $r_{,u}(0,0)=-r_{,v}(0,0)$,  and $A_{,v}(0,0)$, which are determined by the initial mass $M_0$ and charge $Q_0$ of the black hole through the mass and charge function evaluated at the $(0,0)$ location. With this prescription of the initial data, the dynamics of the system can be obtained further by the evolution equations. To test the dependence of our simulations on the initial data, we choose the two sets of initial profiles for our charged scalar field as follows
\begin{eqnarray}
&&1.\quad a=0.02, \quad w=0.5,\quad v_c=3,\nonumber\\
&&2.\quad a=0.03,\quad w=1.0,\quad v_c=3.
\end{eqnarray}

\begin{table}
\begin{ruledtabular}
\begin{tabular}{cccc}
$A$ & $B$ & $C$ & $D$ \\
 $0.213+0.00358i$ & $2.09-0.197i$ & $18.7-0.338i$ & $158-0.666i$
\end{tabular}
\end{ruledtabular}
\caption{The ratio of the dominant quasinormal frequency to the surface gravity at the CH, where the imaginary part in Case D is less than $-1/2$, implying the violation of the SCC at the linear level.}\label{linear}
\end{table}

{\it Mass inflation and strong cosmic censorship}.--As evidenced by
\begin{equation}
M_{,\tau}=QQ_{,\tau}/r-4r^2r_{,u}\Psi_{,\tau}\overline{\Psi}_{,\tau}
\end{equation}
with $\tau=\int  dve^\sigma$ the affine parameter,  the mass inflation near the Cauchy horizon signals the inextendibility beyond the Cauchy horizon of the corresponding weak solution to Einstein equation if  the variation of $Q$ and $r$ are finite over there, as is the case we shall see\footnote{The variation of $r$ is obviously finite because there is no much room for $r$ to decrease due to the fact that the minimal value for $r$ is zero.}.

\begin{figure}
\begin{center}
\includegraphics[width=8cm]{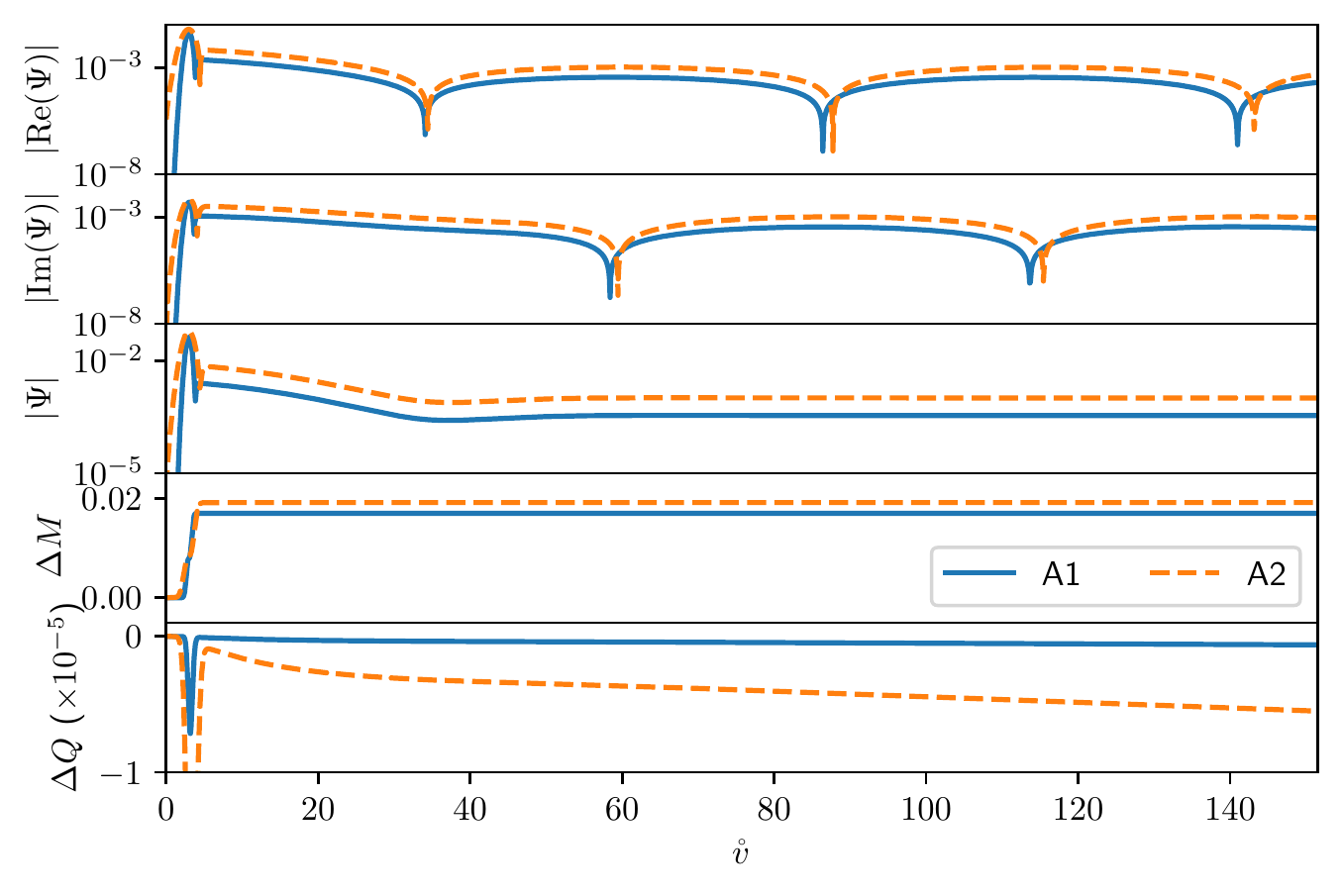}
\includegraphics[width=8cm]{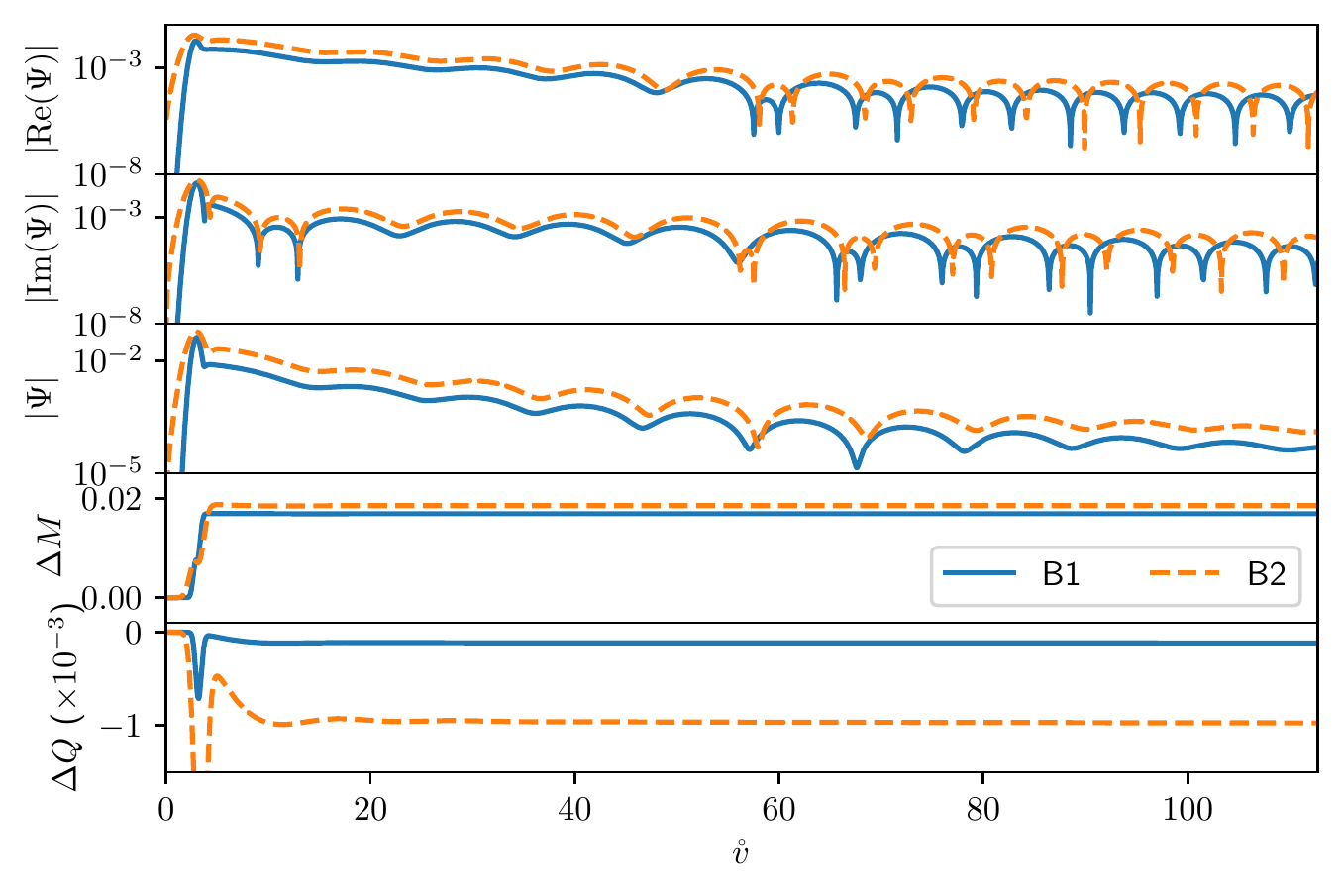}
\includegraphics[width=8cm]{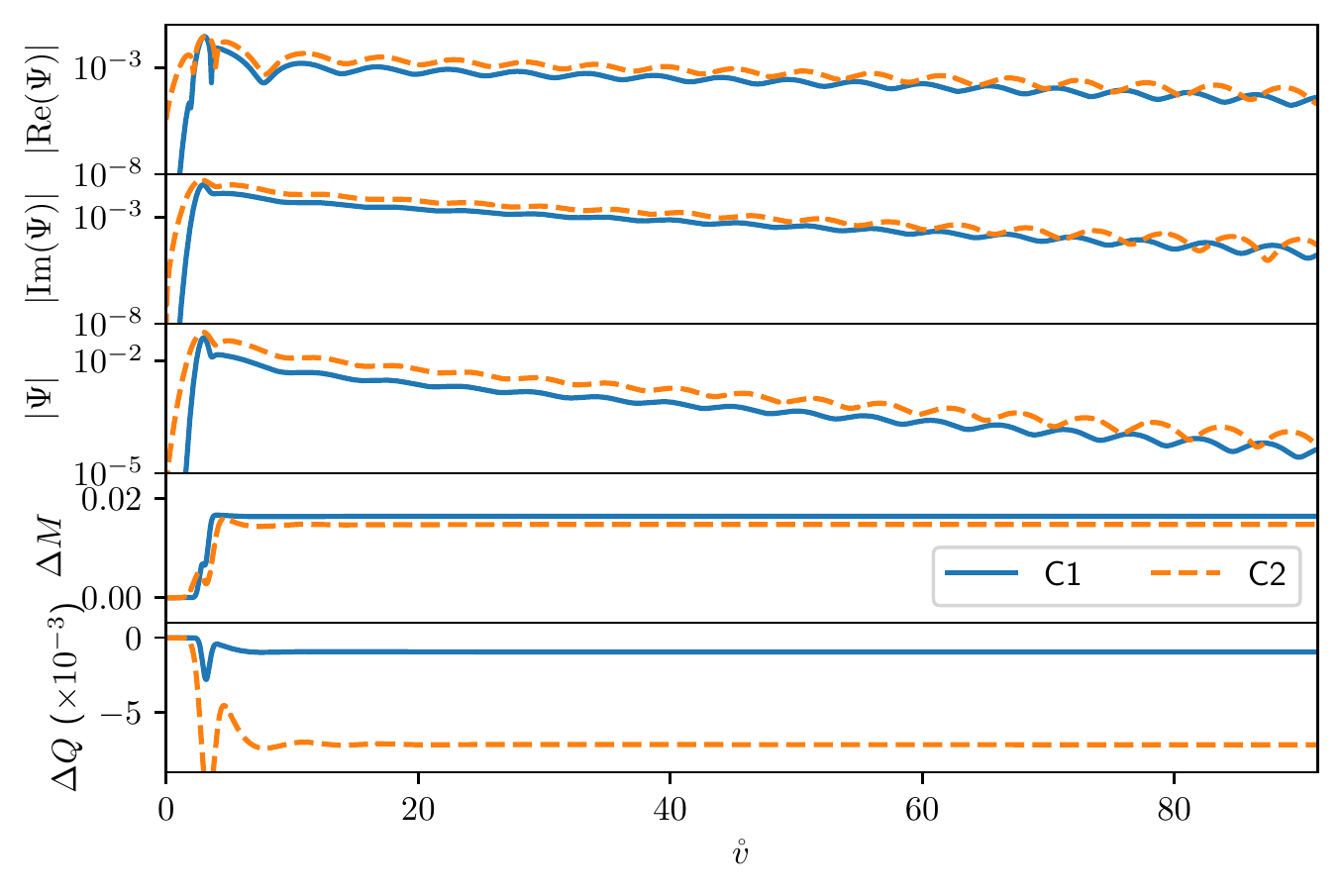}
\end{center}
\caption{The scalar profile, the variation of mass and charge functions along the event horizon with the Eddington time defined as $\mathring{v}=-\int dve^\sigma/(2r_{,u})$ for the case of $M_0=1$, $\Lambda=0.06$, and $Q_0=1.0068$, where A is for $q=0.1$, B for $q=1.0$, and C for $q=2.0$.}
\label{EH}
\end{figure}

\begin{figure}
\begin{center}
\includegraphics[width=8cm]{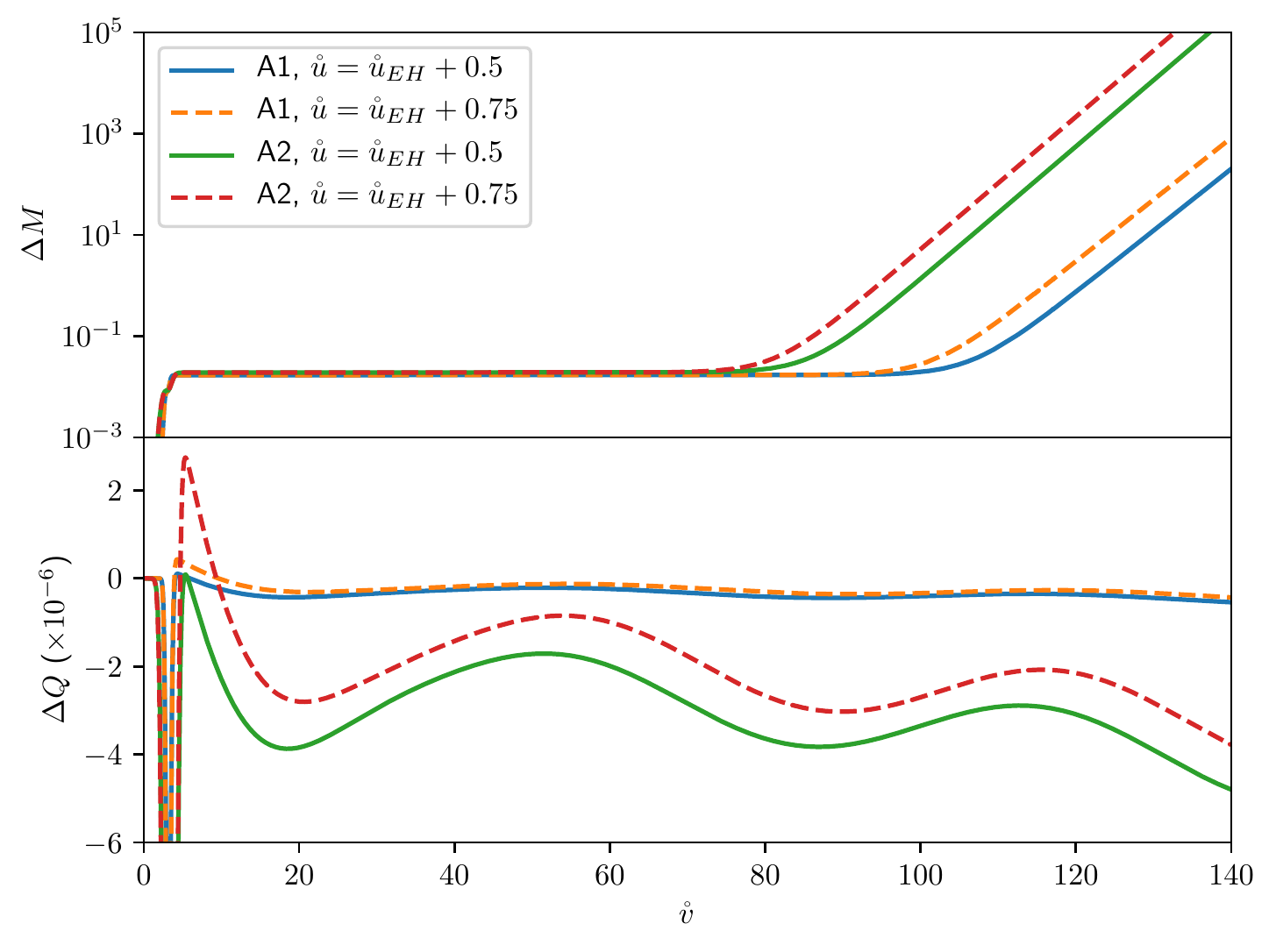}
\includegraphics[width=8cm]{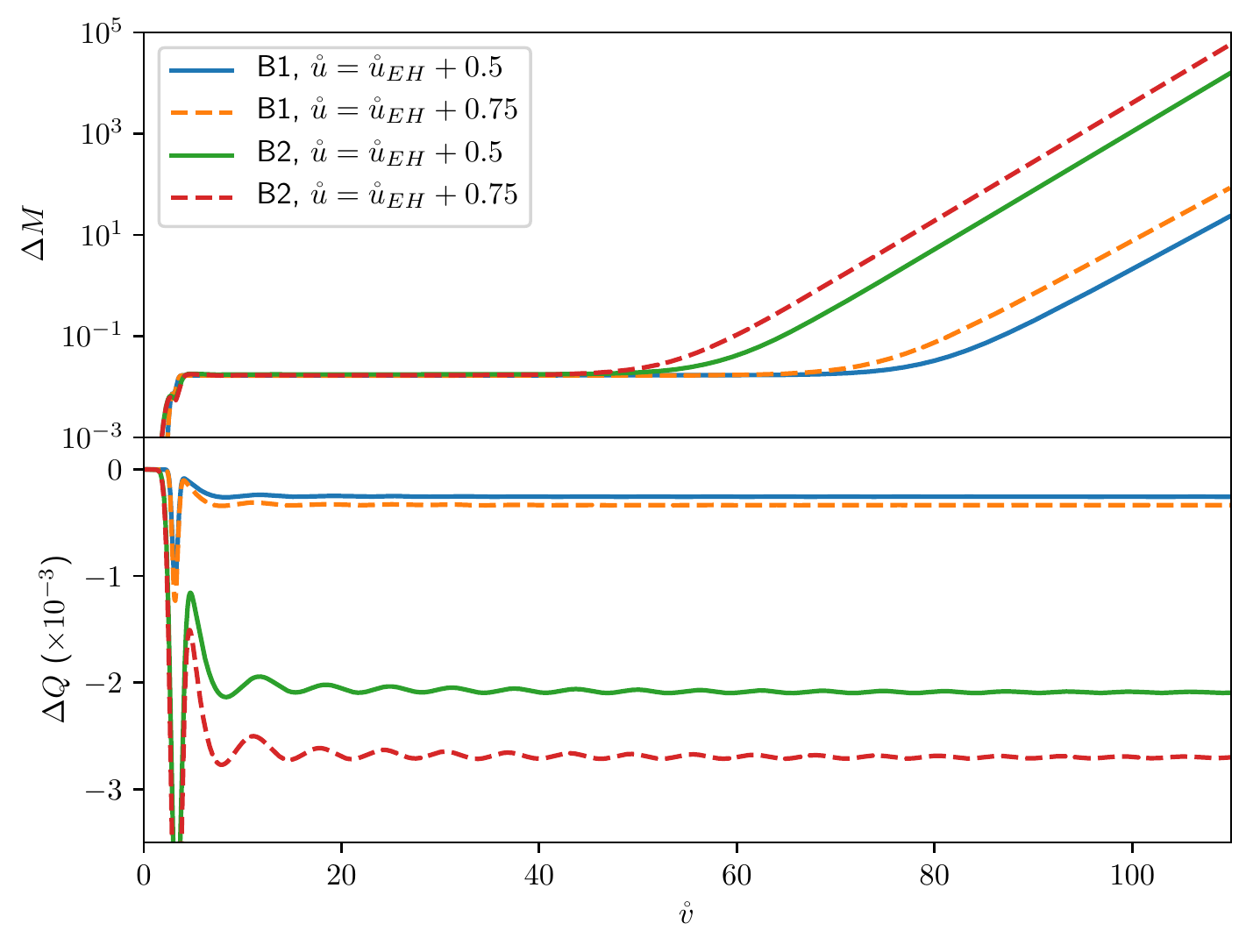}
\includegraphics[width=8cm]{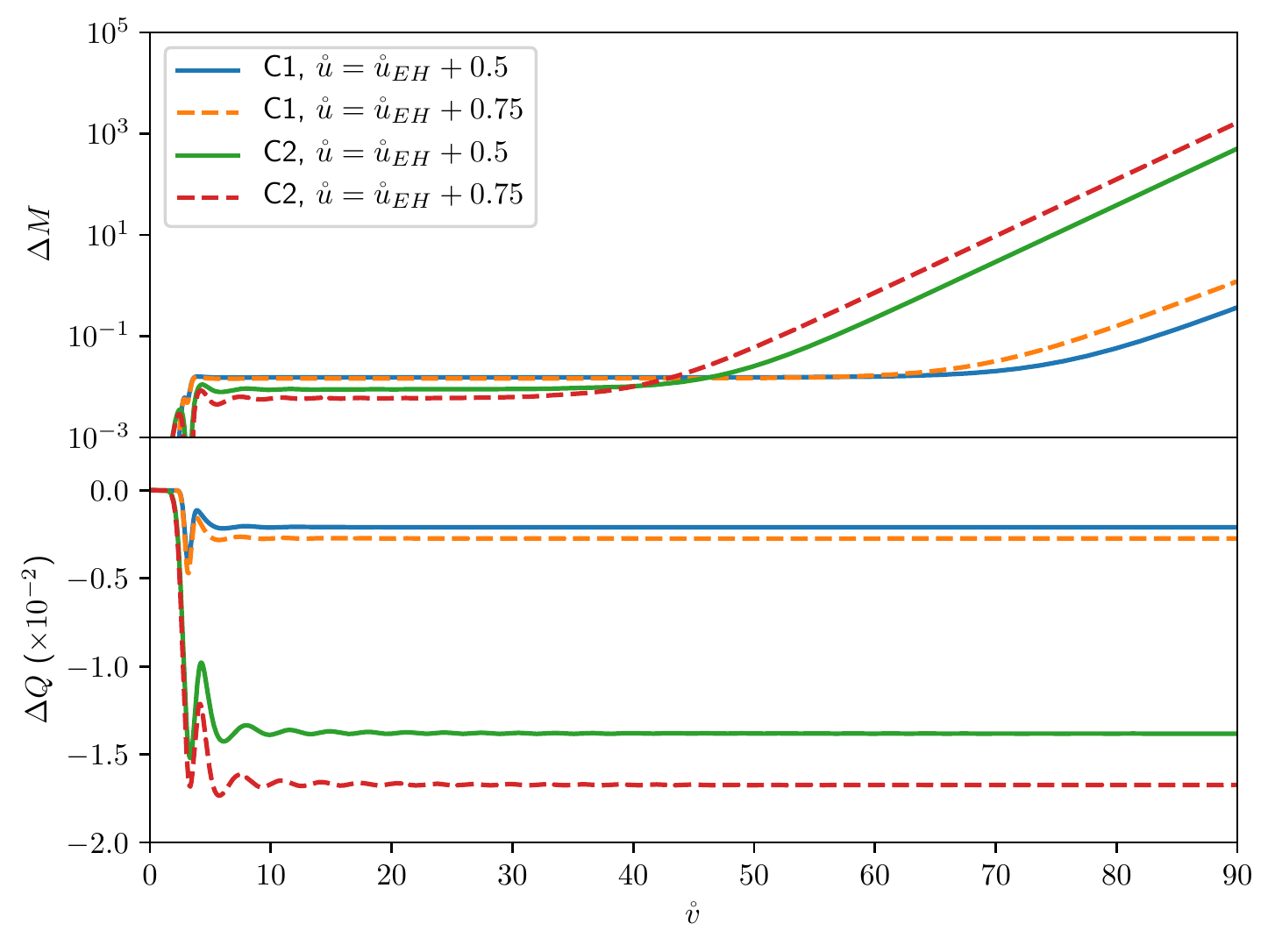}
\end{center}
\caption{The variation of mass and charge functions along two outgoing light curves inside of the black hole event horizon with the affine parameter $u^0$ defined as $\mathring{u}=\int du e^\sigma$ along $v=0$.}
\label{CH}
\end{figure}

We first present the typical result for the case of $M_0=1$, $\Lambda=0.06$, and $Q_0=1.0068$ with $r_0=2.5$ in FIG.\ref{EH} and FIG.\ref{CH}, where the linear result shows no violation of the SCC for the charged perturbations. To be more specific, according to the linear analysis shown in Table \ref{linear}, Case A corresponds to the $q=0.1$ perturbation, which is dominated by the zero superradiant mode, Case B corresponds to the $q=1.0$ perturbation, dominated by the zero oscillatory decaying mode, and Case C corresponds to the $q=2.0$ with the photon sphere mode dominant. The scalar profile in FIG.\ref{EH} demonstrates a consistent picture with such a linear analysis. But as one can see, although the scalar field dies out in both Case B and Case C, the back reaction of the scalar field drives the initial RNdS black hole to another RNdS black hole with the mass increased and the charge decreased as the final state. On the other hand,  due to the tiny growth rate for the zero superradiant mode, it is impossible for us to track the full non-linear evolution towards the final state induced by such a superradiant instability for Case A.  In addition, as depicted in FIG.\ref{CH}, with the well behaved variation of charge in all the cases, the occurrence of the mass inflation
indicates no violation of the SCC also at the full non-linear level, in good agreement with the aforementioned linear result. 
\begin{figure}
\begin{center}
\includegraphics[width=8cm]{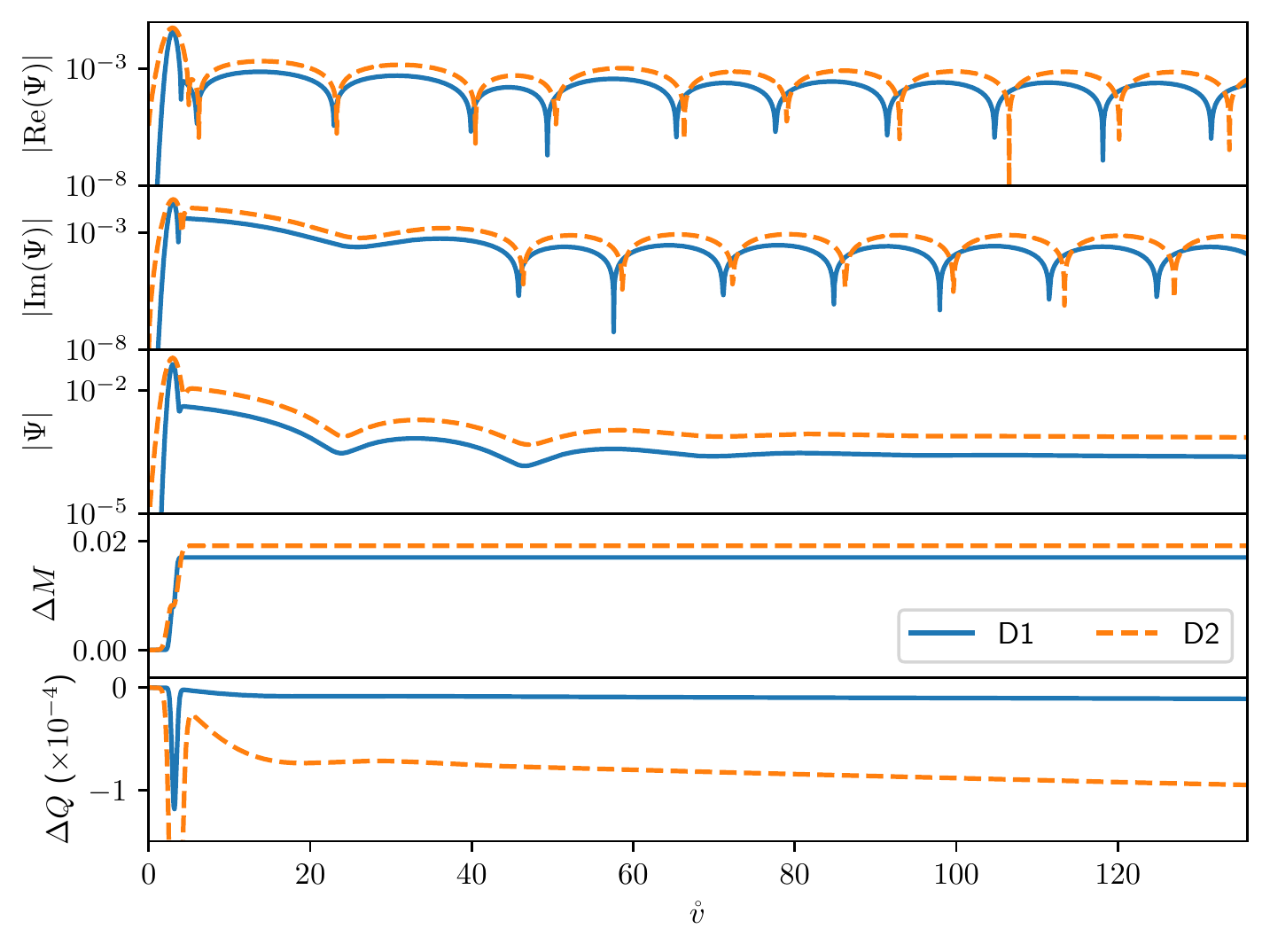}
\includegraphics[width=8cm]{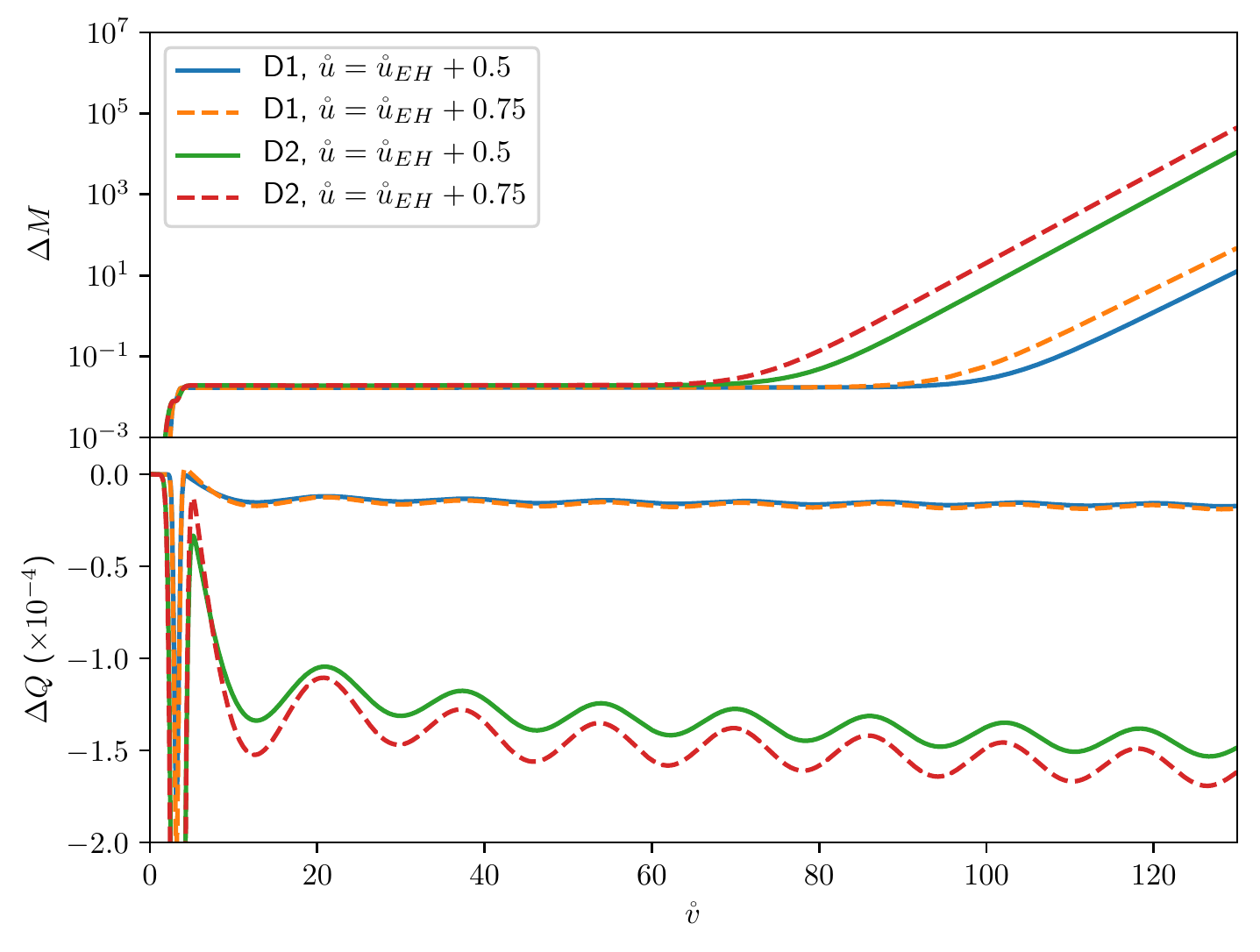}
\end{center}
\caption{The non-linear dynamics for the case of $M_0=1$, $\Lambda=0.06$, $Q_0=1.01085$ , and $q=0.4$.}
\label{No}
\end{figure}

The more important result we next present is the case of $M_0=1$, $\Lambda=0.06$, $Q_0=1.01085$, and $q=0.4$ with $r_0=2.5$. As shown in Table \ref{linear}, the linear result shows the violation of the SCC for this case. But as demonstrated in the bottom panel of FIG.\ref{No}, our non-linear evolution indicates no violation of the SCC, invalidating the linear result. This result seems reasonable since the black hole is driven away from the initial highly near-extremal RNdS black hole by the charged perturbation, which can be seen by the increase of the black hole mass and the decrease of the black hole charge in the top panel of FIG.\ref{No}.

{\it Conclusion}.--
The formation of the RNdS black hole entails the participation of the charged fields. The linear analysis shows that there is still room for the violation of the SCC under the charged scalar perturbation. Taking into account that such a violation regime is so special, we have performed the full non-linear numerical simulation of the charged scalar field minimally coupled with the Einstein-Maxwell system. As a result, we find that the back reaction of the charged perturbation can save the SCC out of the violation regime, which indicates that the SCC keeps as strong as ever. Therefore it is unnecessary to resort to the rough initial data to rescue the SCC.

\begin{acknowledgments}
This work is partially supported by NSFC with Grant No.11675015 and No.11775022 as well as by FWO-Vlaanderen through the project G020714N, G044016N, and G006918N. HZ is supported by the Vrije Universiteit Brussel through the Strategic Research Program ``High-Energy Physics", and he is also an individual FWO Fellow supported by 12G3515N. We like to thank Jorge Santos and Bin Wang for their useful communications at different stages of this project. HZ is grateful to Pau Figueras for his wonderful hospitality at Queen Mary University of London as well as his helpful suggestions. 

\end{acknowledgments}

\section*{Supplemental Material}

Here we present some details on the numerical scheme, adaptive gauges, and error estimates for the simulations of our system. We solve $A$ by the constraint equation (\ref{c3}) while evolve $r$, $\Psi$, $\sigma$ in order through the evolution equations by the fourth-order Runge-Kutta method along the $u$ direction, where along the $v$ direction we employ the Chebyshev pseudospectral multi-domain technique developed in \cite{YS}, which turns out to be highly efficient in decreasing the numerical error at the connection points. We adopt the Eddington-like variant of the horizon resolving gauge\cite{EO}
\begin{equation}
\sigma_1(0,v)=\frac{1}{2}\ln(-4r_{,u}r_{,v}),\quad \sigma_1(u,v_f)=\ln(2r_{,v})+c_1
\end{equation}
where $v_f$ is the end of the computational domain along the $v$ direction and $c_1$ is determined by the continuity of $\sigma$ at the point $(0,v_f)$. This gauge choice can allow us to perform a long time evolution outside of the black hole. But in order to investigate the SCC, we are required to go into the black hole, which can be achieved by the following gauge choice
\begin{eqnarray}
&&\sigma_2(u_f,v)=\sigma_1(u_f,v)+c_2,\nonumber\\
&&\sigma_2(u,0)=\sigma_2(u_f,0)[1-\textrm{erf}(\frac{u-u_f}{b})],
\end{eqnarray}
where $u_f$ is the end of the computational domain for the Eddington-like gauge along the $u$ direction with $c_2$ and $b$ appropriately chosen such as to decrease the numerical error across the black hole. We shall take the constant $u_f$ surface as the proxy for the black hole event horizon.

\begin{figure}
\begin{center}
\includegraphics[width=8cm]{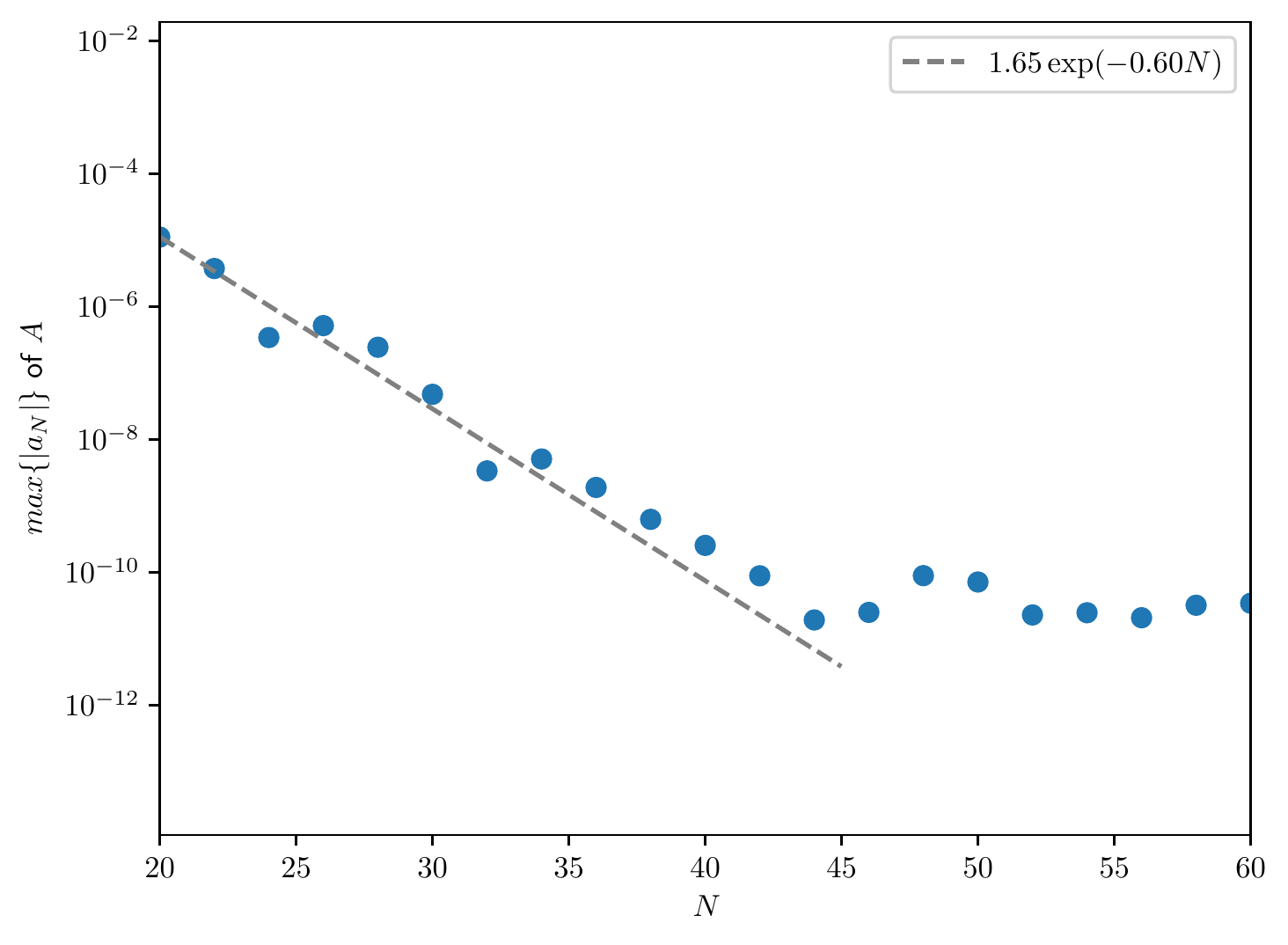}
\end{center}
\caption{The numerical convergence of the maximum of the absolute value of the last spectral coefficient of $A$ at $u_f=100$ for the A1 case, where we have employed $80$ domains in $v\in[0,v_f=200]$ with $N$ points in each domain.}
\label{convergence}
\end{figure}

\begin{figure}
\begin{center}
\includegraphics[width=8cm]{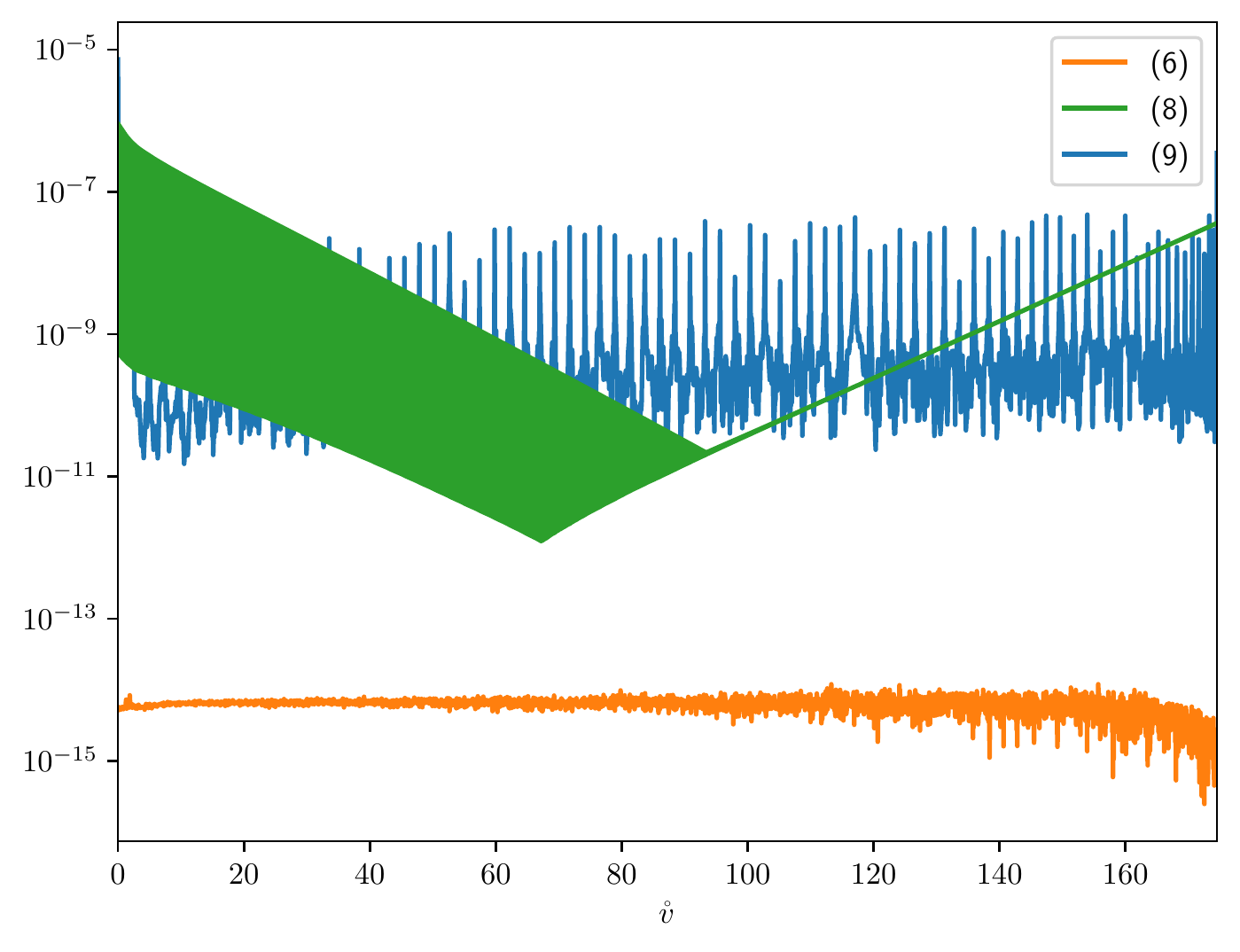}
\end{center}
\caption{The variation of the maximal constraint violations along $u$ with the Eddington time $\mathring{v}$ for the A1 case.}
\label{error}
\end{figure}

We have checked the convergence of our numerical simulation by examining the maximum of the absolute value of the last spectral coefficient over all the domains for each evolution variable. We demonstrate such a convergence for $A$ in FIG.\ref{convergence}, where one can see the exponential convergence up to $N\approx 45$. In addition, since our numerical scheme is a free evolution with Eq.(\ref{Ae}), Eq.(\ref{c1}), and Eq.(\ref{c2}) as the constraints, we have also checked that all of them remain satisfied throughout our evolution. We show the variation of the corresponding maximal constraint violations along $u$ with the Eddington time in FIG.\ref{error} for the A1 case.

\begin{thebibliography}{20}
\bibitem{CCDHJ1}V. Cardoso, J. L. Costa, K. Destounis, P. Hintz, and A. Jansen, Phys. Rev. Lett. 120, 031103(2018).
\bibitem{DRS1}O. J. C. Dias, H. S. Reall, and J. E. Santos, JHEP 1810, 001(2018).
\bibitem{LZCCN}R. Luna, M. Zilhao, V. Cardoso, J. L. Costa, and J. Natario, Phys. Rev. D 99, 064014(2019).
\bibitem{DERS}O. J. C. Dias, F. C. Eperson, H. S. Reall, and J. E. Santos, Phys. Rev. D 97, 104060(2018).
\bibitem{Hod1}S. Hod, Phys. Lett. B 780, 221(2018).
 \bibitem{Hod2}S. Hod, arXiv:1810.00886.
 \bibitem{Hod3}S. Hod, Nucl. Phys. B 941, 636(2019).
 \bibitem{CCDHJ2}V. Cardoso, J. L. Costa, K. Destounis, P. Hintz, and A. Jansen, Phys. Rev. D 98, 104007(2018)
 \bibitem{MTWZZ}Y. Mo, Y. Tian, B. Wang, H. Zhang, and Z. Zhong, Phys. Rev. D 98, 124025(2018)
 \bibitem{DRS2}O. J. C. Dias, H. S. Reall, and J. E. Santos, Class. Quant. Grav. 36, 045005(2019).
 \bibitem{DS}M. Dafermos and Y. Shlapentokh-Rothman, Class. Quant. Grav. 35, 195010(2018).


\bibitem{YS}H. H. Yang and B. Shizgal, Comput. Methods Appl. Mech. Engrg. 118, 47(1994).
 \bibitem{EO}E. Eilon and A. Ori, Phys. Rev. D 93, 024016(2016).

 








\end{thebibliography}
\end{document}